\documentclass{revtex4}

\usepackage[dvips]{graphicx}
\usepackage{xr}

\usepackage{amssymb,amsfonts,amsmath,float}

\newlength{\colw}
\newfloat{algorithm}{tbp}{loa}
\floatname{algorithm}{Algorithm}

\begin{document}


\title{Efficient supervised learning in networks with binary synapses}

\author{Carlo Baldassi}
\affiliation {ISI Foundation, Viale S. Severo 65,
    I-10133 Torino, Italy}
\author{Alfredo Braunstein}
\affiliation{Politecnico di Torino, C.so Duca
    degli Abruzzi 24, I-10129 Torino, Italy}
\affiliation {ISI Foundation, Viale S. Severo 65,
    I-10133 Torino, Italy}
\author{Nicolas Brunel}
 \affiliation{Laboratory of Neurophysics and Physiology (UMR
    8119), CNRS-Universit\'{e} Paris 5 Ren\'{e} Descartes, 45 rue des
    Saints P\`{e}res, 75270 Paris Cedex 06}
\affiliation {ISI Foundation, Viale S. Severo 65,
    I-10133 Torino, Italy}
\author{Riccardo Zecchina}
\affiliation{Politecnico di Torino, C.so Duca
    degli Abruzzi 24, I-10129 Torino, Italy}
\affiliation{ICTP, Strada Costiera 11, I-34100
    Trieste, Italy}



\begin{abstract}
Recent experimental studies indicate that synaptic changes induced by
neuronal activity are discrete jumps between a small number of stable
states. Learning in systems with discrete synapses is known to be a
computationally hard problem. ÊHere, we study a neurobiologically
plausible on-line learning algorithm that derives from Belief Propagation algorithms. 
We show that it performs remarkably well in a model neuron with binary synapses, and 
a finite number of `hidden' states per synapse, that has to learn
a random classification task. Such system is able to learn a number
of associations close to the theoretical limit, in time which is
sublinear in system size.  This is to our knowledge the first on-line algorithm that is able 
to achieve efficiently  a finite number of patterns learned per binary synapse.  Furthermore, 
we show that performance is optimal for a finite number of hidden  states which becomes 
very small for sparse coding. 
The algorithm is similar to the standard `perceptron' learning algorithm, with an
additional rule for synaptic transitions which occur only if a
currently presented pattern is `barely correct'. In this case, the
synaptic changes are meta-plastic only (change in hidden states and
not in actual synaptic state), stabilizing the synapse in its current
state. ÊFinally, we show that a system with two visible states and $K$
hidden states is much more robust to noise than a system with $K$
visible states. ÊWe suggest this rule is sufficiently simple to be
easily implemented by neurobiological systems or in hardware.
\end{abstract}
\maketitle

\section{Introduction}

Learning and memory are widely believed to occur through
mechanisms of synaptic plasticity. In spite of a huge amount of
experimental data documenting various forms of plasticity, as
e.g.~long-term potentiation (LTP) and long-term depression (LTD), the
mechanisms by which a synapse changes its efficacy, and those by which
it can maintain these changes over time remain unclear. Recent
experiments have suggested single synapses could be similar to noisy
binary switches \cite{petersen98,oconnor05}.  Bistability could be in
principle induced by positive feedback loops in protein interaction
networks of the post-synaptic density
\cite{lisman85,zhabotinsky00,bhalla99}.  Binary synapses would have
the advantage of robustness to noise and hence could preserve memory
over long time scales, compared to analog systems which are typically
much more sensitive to noise.

Many neural network models of memory use binary synapses to store
information
\cite{willshaw69,marr69,sompolinsky86,amit92b,amit94b,brunel98b,fusi05}.
In some of these network models, learning occurs in an unsupervised
way.  From the point of view of a single synapse, this means that
transitions between the two synaptic states (a state of low or zero
efficacy, and a state of high efficacy) are induced by pre and
post-synaptic activity alone. Tsodyks \cite{tsodyks90} and Amit and
Fusi \cite{amit92b,amit94b} have shown that the performance of such
systems (in terms of information stored per synapse) is very poor,
unless two conditions are met: (1) activity in the network is sparse
(very low fraction of neurons active at a given time); and (2)
transitions are stochastic, with in average a balance between up
(LTP-like) and down (LTD-like) transitions. This poor performance has
motivated further studies \cite{fusi05} in which hidden states are
added to the synapse in order to provide it with a multiplicity of
time scales, allowing for both fast learning and slow forgetting.

In a supervised learning scenario, synaptic modifications are induced
not only by the activity of pre and post-synaptic neurons but also by
an additional `teacher' or `error' signal which gates the synaptic
modifications. The prototypical network in which this type of learning
has been studied is the one-layer perceptron which has to perform a
set of input-output associations, i.e.~learn to classify correctly
input patterns in two classes. In the case of analog synapses,
algorithms are known to converge to synaptic weights that solve the
task, provided such weights exist \cite{rosenblatt62,minsky69}. On the
other hand, no efficient algorithms are known to exist in a perceptron
with binary (or more generally with a finite number of states)
synapses, in the case the number of patterns to be learned scales with
the number of synapses.  In fact, studies on the capacity of binary
perceptrons \cite{krauth89,krauth89b} used complete enumeration
schemes in order to determine numerically the capacity.  These studies
found a capacity of about 0.83 bits per synapse in the random input --
output categorization task, very close to the theoretical upper bound
of 1 bit per synapse. However, it is not even clear whether there
exist efficient algorithms that can reach a finite capacity per
synapse, in the limit of a large network size $N$. Indeed, learning in
such systems is known to be a NP-complete task
\cite{NPC-learning1,NPC-learning2}.

Recently, `message passing' algorithms have been devised that solve
efficiently non-trivial random instances of NP-complete optimization
problems, like e.g. K-satisfiability or graph
coloring~\cite{MPZ,MZ,BMZ,COLORING}. One such algorithm, Belief
Propagation (BP), has been applied to the binary perceptron problem
and has been shown to be able to find efficiently synaptic weight
vectors that solve the classification problem for a number of patterns
close to the maximal capacity (above 0.7 bits per
synapse)\cite{nosotros}.  However, this algorithm has a number of
biologically unrealistic features (e.g.~memory stored in several
analog variables). Here, we explore algorithms that are inspired from
the BP algorithm but are modified in order to make them biologically
realistic.

The paper is organized as follows: First we present the general scheme
for the simplest setup of $\pm1$ patterns and synapses as well as
results with bounded and unbounded hidden variables. Then we discuss
the more realistic 0,1 case with results including the sparse coding
limit.  Implications of our results are discussed in the concluding
section. Details are given in the Supporting Information.


\section{Binary $\pm$ 1 neurons and synapses}
\label{sec:pm1}
\subsection{The model neuron}

We consider a neuron with two states (`inactive' and `active')
together with its $N$ presynaptic inputs which we take to be also
binary.  Depending on the time scale, these two states could
correspond in a biological neuron either to emission of a single spike
or not, or to elevated persistent activity or not.  The strength of
synaptic weights from presynaptic neuron $i$ ($i=1,\ldots,N$) is
denoted by $w_{i}$. Given an input pattern of activity
$\{\xi_{i},i=1,\ldots,N\}$, the total synaptic input received by the
neuron is $I=\sum_{i=1}^{N}w_{i}\xi_{i}$.  The neuron is active if
this total input is larger than a threshold $\theta$, and is inactive
otherwise. Such a model neuron is sometimes called a \emph{perceptron}
\cite{rosenblatt62}. In this paper we consider binary synaptic
weights. In addition, each synapse is characterized by a discrete
`hidden variable' that determines the value of the synaptic weight. In
this section we consider $\{-1,+1\}$ neurons and synapses, and
$\theta=0$; in order to simplify the notation, we will also assume
$N$ to be odd, so that the total synaptic input is never equal to $0$.
This assumption can be dropped when dealing with $\{0,1\}$ model neurons.

\subsection{The classification problem}

We assume that our model neuron has to classify a set of $p=\alpha N$
random input patterns $\{\xi_{i}^{a},i=1,\ldots,N,a=1,\ldots,p\}$
into two classes (active or inactive neuron, $\sigma^{a}=\pm1$).
The set of patterns which should be classified as $+1$ ($-1$) is
denoted by $\Xi_{+}$ ($\Xi_{-}$) respectively. In each pattern,
the activity of input neurons is set to 1 or $-1$ with probability 0.5,
independently from neuron to neuron and pattern to pattern. The learning
process consists in finding a vector of synaptic weights $\mathbf{w}$
such that \emph{all} patterns in $\Xi_{+}$(resp. $\Xi_{-}$) are
mapped onto output $+$ (resp. $-$). Hence, the vector $\mathbf{w}$
has to satisfy the $p$ equations 
\begin{equation}
\sigma^{a}=\mbox{sign}\left(\sum_{j=1}^{N}w_{j}\xi_{j}^{a}\right)\quad\mbox{for $a=1,\ldots,p$}.
\end{equation}
 We will call such a vector a \emph{perfect classifier} for this problem.

In the case of $\pm1$ synapses and inputs we are considering in this
section, the problem is the same if we consider the set $\Xi_{-}$ to
be empty, i.e. $\sigma^{a}=+1$ for all $a=1,\dots,p$, as we can
always redefine $\xi_{j}^{a}\to\sigma_{j}^{a}\xi_{j}^{a}$ and require
the output to be positive. This will no longer hold in next section.

\subsection{The perceptron learning algorithm}

In the case of unbounded synaptic weights, there exists a standard
learning algorithm that can find a perfect classifier, provided such
a classifier exists, namely the perceptron algorithm (\textbf{SP})
\cite{rosenblatt62,minsky69}. The algorithm consists in presenting
sequentially the input patterns.  When at time $\tau$ pattern
$\xi^\tau$ is presented, one first computes the total input
$I=\sum_{i=1}^{N}w^{\tau}_{i}\xi_{i}^{\tau}$ and then:

{\sf
\begin{itemize}
\item If $I\geq0$: do nothing. 
\item If $I<0$: change the synaptic weights as follows: \[
w^{\tau+1}_{i}=w^{\tau}_{i}+\xi^{\tau}_{i}\]
\end{itemize}
}

This algorithm has the nice feature that it is guaranteed to converge
in a finite time if a solution to the classification problem exists.
Furthermore, it has other appealing features that makes it a plausible
candidate for a neurobiological system: the only information needed to
update the synapse is an `error signal' (the synapse is modified only
when the neuron gave an incorrect output), and the current activity of
both presynaptic and postsynaptic neurons. However, the convergence
proof exists for unbounded synaptic weights \cite{rosenblatt62}, or
for sign-constrained synaptic weights \cite{amit89}, but not when
synaptic weight can take only a finite number of states.

\subsection{Requirements for biologically plausible learning
  algorithms with binary weights}

In this paper, we explore learning algorithms for binary synaptic
weights. Each synapse is endowed with an additional discrete `hidden'
variable $h_{i}$. This hidden variable could correspond to the state
of the protein interaction network of the post-synaptic density, which
can in principle be multistable due to positive feedback loops in such
networks \cite{bhalla99,zhabotinsky00}. Each synaptic weight $w_{j}$
will depend solely on the sign of the corresponding hidden variable
$h_{j}$; in the following, in order to avoid the ambiguous $h_i=0$
state, we will always represent the hidden variables by odd integer
numbers (this simplifies the notation but doesn't affect the
performance).  We first consider the (unrealistic) situation of an
unbounded hidden variable, and then investigate learning with bounded
hidden variables.  Similar to the perceptron algorithm, we seek
`on-line' algorithms (i.e. modifications are made only on the basis of
the currently presented pattern) which, at each time step $\tau$,
modify synapses based only on variables available to a synapse:
\textbf{(i)}~The current total synaptic input $I^{\tau}$ and hence the
current post-synaptic activity; \textbf{(ii)}~The current presynaptic
activity $\xi_{i}^{\tau}$; \textbf{(iii)}~An error signal indicating
whether the output was correct or not.
At each time step, the current input pattern is drawn randomly from
the set of patterns, and the hidden variables $h_{j}^{\tau}\to
h_{j}^{\tau+1}$ and the synaptic weights $w_{j}^{\tau}\to
w_{j}^{\tau+1}$ are updated according to the algorithm.

\subsection{Quantifying performance of various algorithms}

The maximal number of patterns for which a weight vector can be found
is $\alpha_{\max}\simeq 0.83$ for random unbiased patterns
\cite{krauth89}.  Hence, the performance of an algorithm can be
quantified by how close the maximal value of $\alpha$ at which it can
find a solution is to $\alpha_{\max}$. In practice, one has to
introduce a maximal number of iterations per pattern. For example, a
complete enumeration algorithm (in which one checks sequentially the
$2^{N}$ possible synaptic weight configurations) is guaranteed to find
a solution for any $\alpha\leq\alpha_{\max}$, but it finds it in an
implausibly long time (exponentially large in $N$). Here, we impose a
maximal number of iterations (typically $10^4$ per pattern) and find
the maximal value of $\alpha$ for which a given algorithm is able to
find a solution.

\subsection{Belief propagation--inspired algorithms}
A modification of the Belief Propagation (BP) algorithm was found by
Braunstein and Zecchina\cite{nosotros} to perform remarkably well in
the random binary perceptron problem.  However, the BP algorithm has
some features which make it unplausible from the biological point of
view.  In the Supporting Information, we show that with a number of
simplifications, this algorithm can be transformed into a much simpler
on-line one that satisfies all the requirements outlined above. The
resulting algorithm is as follows:

\subsubsection*{BP-inspired (BPI) Algorithm}

{\sf Compute $I=\xi^{\tau}\cdot w^{\tau}$, where $w_i^{\tau}=\textrm{sign}\left(h_i^{\tau}\right)$, then
\begin{enumerate}

\item [(R1)]If $I>1$, do nothing
\item [(R2)]If $I=1$ then:
\begin{enumerate}
\item If $h_{i}^{\tau}\xi_{i}^{\tau}\geq1$, then $h_{i}^{\tau+1}=h_{i}^{\tau}+2\xi_{i}^{\tau}$
\item Else do nothing
\end{enumerate}
\item [(R3)]If $I\leq-1$ then $h_{i}^{\tau+1}=h_{i}^{\tau}+2\xi_{i}^{\tau}$.
\end{enumerate}}

These rules can be interpreted as follows. (R1) As $I>1$ the synaptic
input is sufficiently above threshold, such that a single synaptic (or
single neuron) flip would not affect the neuronal output; therefore
all variables are kept unchanged. (R2) As $I=1$ the synaptic input is
just above threshold (a single synaptic or single neuron flip could
have potentially brought it below threshold), then some of the hidden
variables need to be changed. The variables that are changed are those
that were going in the right direction, i.e. those that contributed to
having the synaptic input go above threshold. Finally for (R3) $I<0$
so the output is incorrect and then all hidden variables need to be
changed.  The factor of $2$ included in rules R2 and R3 guarantees
that the hidden variables will still be odd when updated if they are
initialized to be so.

Note that this algorithm has two distinct features compared to the
perceptron algorithm:
\textbf{(i)} Hidden variables obey update rules that are similar to those of
the \textbf{SP} algorithm, but the actual synaptic weight is binary;
\textbf{(ii)} One of the update rules, rule R2 (corresponding to a synaptic input
just above threshold), is new compared to \textbf{SP}. 

To investigate the effect of rule R2 on performance, we also simulated
a stochastic version of the \textbf{BPI} algorithm, in which such a
rule is only applied with probability $p_{s}$ for each presented pattern:

\subsubsection*{Stochastic BP-inspired (SBPI) Algorithm}

As \textbf{BPI}, except rule R2 is replaced by:
{\sf
\begin{enumerate}
\item[(R2)] If $I=1$, then:
\begin{enumerate}
\item with probability $p_{s}$:
\begin{enumerate}
\item If $h_{i}^{\tau}\xi_{i}^{\tau}\geq1$, then $h_{i}^{\tau+1}=h_{i}^{\tau}+2\xi_{i}^{\tau}$
\item Else do nothing
\end{enumerate}
\item with probability $1-p_{s}$, do nothing
\end{enumerate}
\end{enumerate}
}

\begin{figure}
\begin{center}\begin{tabular}{|c|c|c|c|c|c|}
\cline{1-1} \cline{3-4} \cline{6-6} 
\multicolumn{1}{|c|}{CP}&
&
\multicolumn{2}{|c|}{BPI}&
&
\multicolumn{1}{c|}{Cascade}
\\
\cline{1-1} \cline{3-4} \cline{6-6} 
$\xi\cdot\mathbf{w}<0$&
&
$\xi\cdot\mathbf{w}<0$&
$\xi\cdot\mathbf{w}=1$&
&
$\xi\cdot\mathbf{w}<0$
\\
\cline{1-1} \cline{3-4} \cline{6-6}
&
&
&
&
&
\\
\includegraphics[scale=0.7]{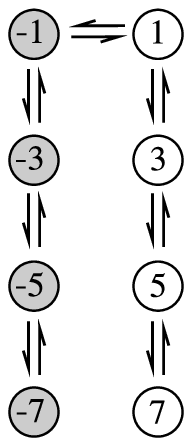}&
&
\includegraphics[scale=0.7]{fig1-sta1.eps}&
\includegraphics[scale=0.7]{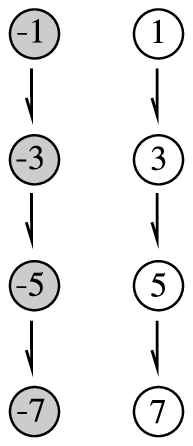}&
&
\includegraphics[scale=0.7]{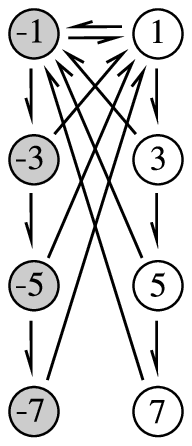}
\\
\cline{1-1} \cline{3-4} \cline{6-6} 
\end{tabular}\end{center}

\caption{
Schematic representation of transitions between synaptic
states in the \textbf{CP} algorithm and the \textbf{BPI} algorithm.
The cascade model introduced by Fusi et al \cite{fusi05} is shown for
comparison.  Circles represent the possible states of the internal
synaptic variable $h_{i}$. Grey circles correspond to $w_i=-1$, white
ones to $w_i=1$.  Clockwise transitions happen when $\xi_{i}=1$,
counter-clockwise when $\xi_{i}=-1$.  Horizontal transitions are
plastic (change value of synaptic efficacy $w_i$), vertical ones
meta-plastic (change internal state only). Downwards transitions make
the synapse less plastic, upward ones more plastic. When the output of
the neuron is erroneous, $\xi\cdot w<0$: transitions occur to the
nearest neighbor internal state.  In the \textbf{CP} algorithm, when
the output is correct, $\xi\cdot w>0$: no transitions occur.  In the
\textbf{BPI} algorithm, when the output is barely correct $\xi\cdot
w=1$ (a single synaptic flip could have caused an error): transitions
are made towards less plastic states only. When the output is safely
correct, $\xi\cdot w>1$: no transitions occur. In the cascade model,
`down' transitions are towards nearest neighbors, while `up' transitions
are towards the highest state with opposite sign. Transition probabilities
decrease with increasing $|h|$, see\cite{fusi05} for more details
\label{fig:algos}
}
\end{figure}

When the parameter \textbf{$p_{s}$} is set to \textbf{$1$}, one
recovers the deterministic \textbf{BPI} algorithm, while setting it to \textbf{$0$}
(thus, in fact, removing rule R2) transforms it into a
`clipped perceptron algorithm' (\textbf{CP}), i.e. a perceptron
algorithm but with clipped synaptic weights.) Both \textbf{BPI} and
\textbf{CP} algorithms are sketched in Fig.~\ref{fig:algos}.

\subsection{Performance with unbounded variables}

\begin{figure}
\begin{center}  
\includegraphics[angle=270,width=1\columnwidth]{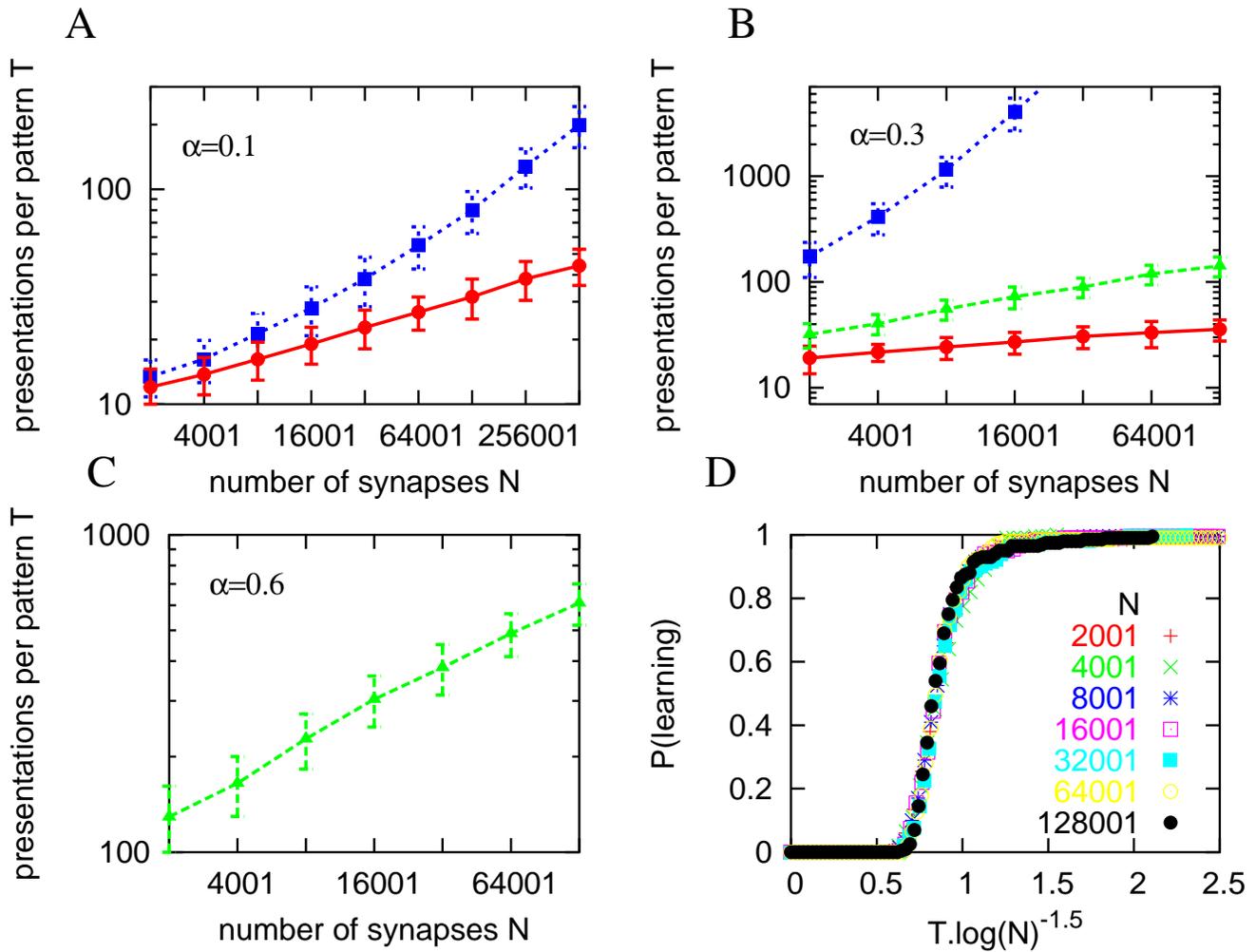}
\end{center}
\caption{Performance of the \textbf{BPI} algorithm with unbounded
hidden variables: \textbf{A-C} convergence time vs. $N$ for different
values and $\alpha$ (indicated on each graph). Points correspond to
number of iterations per pattern until the algorithm converges
averaged over 200 pattern sets, vertical bars are standard
deviations. Dotted lines: \textbf{CP}, Solid lines: \textbf{BPI},
Dashed lines: \textbf{SBPI} with $p_{s} = 0.3$.  The latter is the
only one which can reach $\alpha=0.6$, but performs worse than
\textbf{BPI} for $\alpha \le 0.3$ (it is absent from panel \textbf{A} for clarity).
\label{fig:both}
\textbf{D.}
Probability that the \textbf{BPI} algorithm learns
perfectly $0.3\cdot N$ patterns in less than $T= x\cdot \log(N)^{1.5}$
iterations per pattern vs $x$ for various values of $N$
\label{fig:logscaling}
}
\end{figure}

The performance of both deterministic and stochastic versions of the
\textbf{BPI} algorithm was first investigated numerically with
unbounded hidden variables, for different values of $\alpha$, $N$ and
$p_{s}$.  It turns out that \textbf{SBPI} performs remarkably well,
provided the probability $p_s$ is chosen appropriately - with
$p_s\approx 0.3$ the system can reach a capacity of order 0.65 with a
convergence time that increases with $N$ in a sub-linear fashion (see
Fig.~\ref{fig:both}). On the other hand, the deterministic
\textbf{BPI} ($p_s=1$) has a significantly lower capacity
($\alpha\approx 0.3$), but for those lower values of $\alpha$ it
performs significantly faster than the \textbf{SBPI} algorithm - for
$\alpha=0.3$ the time increases approximately as $(\log N)^{1.5}$, as
shown in Fig.~\ref{fig:logscaling}D.  As an example, the algorithm
perfectly classifies $38400$ patterns with $128001$ synapses with
around $35$ presentations of each pattern only.  By eliminating
completely rule R2 (i.e. \textbf{CP}) convergence time becomes
exponential in $N$ rather than logarithmic, for every tested value of
$\alpha$, as shown by the supralinearity of the blue curves in
Fig.~\ref{fig:both}. Hence, the specificity of rule R2 with respect to
synapses (only synapses that actually went in the right direction for
the current pattern should be modified) is a crucial feature which
makes the \textbf{BPI} algorithm qualitatively superior.  Moreover the
convergence time increases only mildly with $\alpha$, as shown in
Fig.~\ref{fig:both}.

We also find that there is a tradeoff between convergence speed and
capacity: for each value of $\alpha$, there is an optimal value of
$p_s$ that minimizes average convergence time
(shown in the Supporting Information, Fig.~6).
This optimal value decreases with $\alpha$; for $\alpha=0.3$ it is
close to 1, and decreases to 0.3 at $\alpha=0.65$.  Hence, decreasing
$p_s$ enhances the capacity, at the cost of a slower convergence;
nevertheless Fig.~\ref{fig:both}C shows that for values of $\alpha\le 0.60$
\textbf{SBPI} ($p_s=0.3$) learns perfectly the set of input/output
associations in a time that scales sub-linearly with $N$.  Above
$\alpha\geq 0.7$ the algorithm fails to solve instances in a time
shorter than the chosen cutoff time of $10^{4}$. Note that for
$p_s=0.3$ the convergence time depends in a more pronounced way on
$\alpha$ than in the $p_s=1$ case.

We have also investigated an algorithm in which $p_s$ is itself a
dynamical variable that depends on the fraction of errors averaged
over a long time window - such an algorithm with an adaptive $p_s$ is
able to combine faster convergence at low values of $\alpha$ with high
capacity associated with low values of $p_s$ (not shown).


\subsection{Performance with bounded hidden variables}
We now turn to the situation when there is only a limited number of
states $K$ of the hidden variables $h_i$, since it is unrealistic to
assume that a single synapse can maintain an arbitrarily large number
of hidden states.  Thus, we investigated the performance of an
algorithm with symmetrical hard bounds on the values of the hidden
states, $\left|h_i\right| \le K-1$ for all $i$.

\begin{figure}
\begin{center}
\vspace{-0.5cm}
\begin{center}  
\includegraphics[width=1\columnwidth]{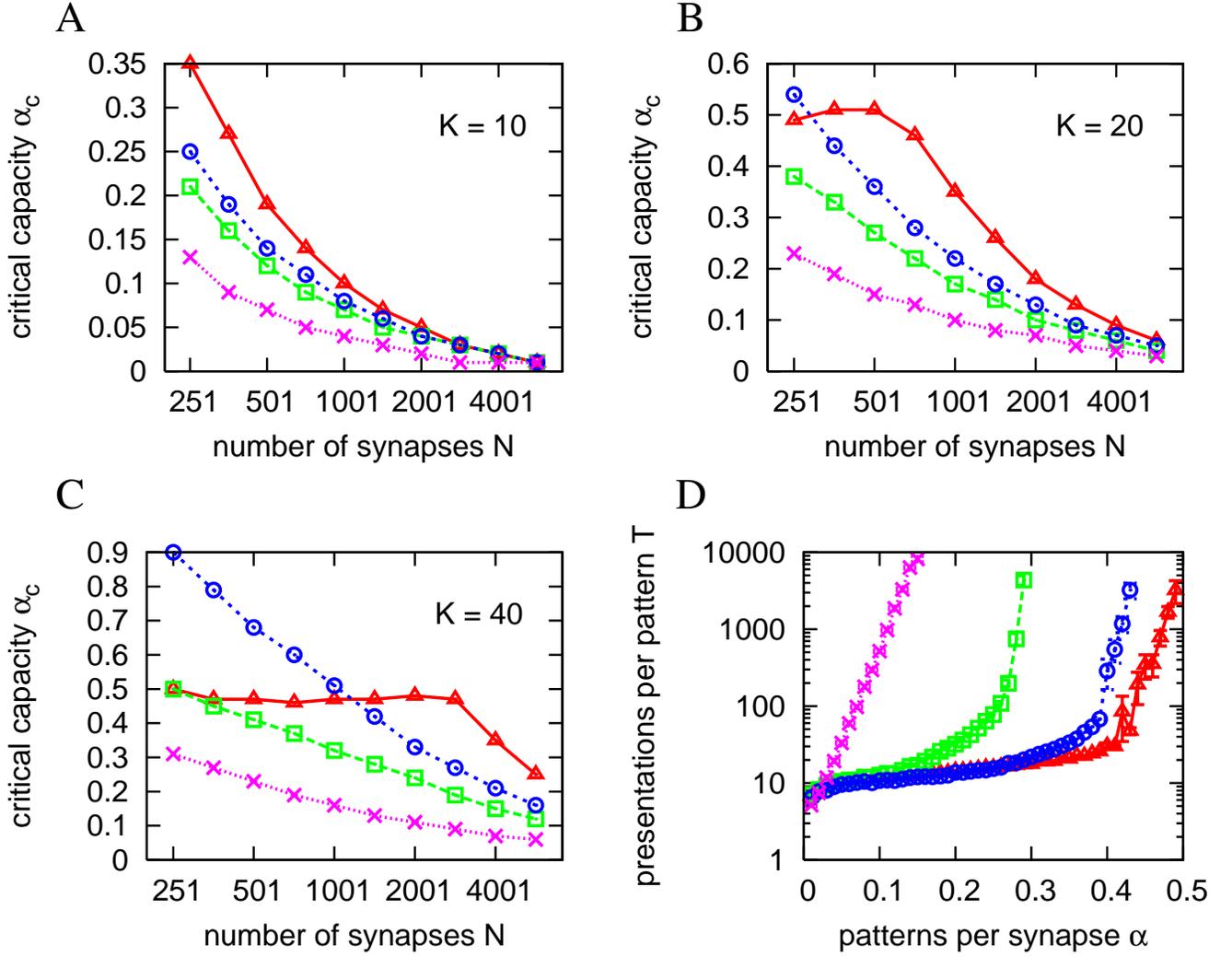}
\end{center}
\end{center}
\caption{Performance of various algorithms with hard-bounded hidden
variables. Triangles: \textbf{BPI}, squares: \textbf{CP},
circles: \textbf{SP}, crosses: cascade model.
\textbf{A}, \textbf{B}, \textbf{C}. Critical capacity vs $N$, with fixed number of
internal states $K$.
\label{fig:ac_vs_n}
\textbf{D}. Convergence time vs $\alpha$
at $N=1415$, $K=40$, averaged over $100$ samples. Figures for
different number of states and synapses are qualitatively similar
\label{fig:iter_cfr}
}
\end{figure}

Figure~\ref{fig:ac_vs_n} shows what happens when the number of
internal states is kept fixed while varying $N$.  For the number of
states we have considered, ($10\le K \le 40$), the optimal value of
$p_s$ is $1$, since in general the stochastic version of the algorithm
requires a larger number of states to be efficient.  Here, we defined
the capacity as the number of patterns for which there is $90\%$
probability of perfect learning in $10^{4}$ iterations, and plotted in
Fig.~\ref{fig:ac_vs_n} the corresponding critical $\alpha$ against $N$
for different values of the states number $K$, comparing \textbf{BPI},
\textbf{CP}, and the cascade model (defined as in Fig.~\ref{fig:algos}).
We also compared these algorithms that have only 2 `visible' synaptic
states but $K$ hidden states, with the \textbf{SP} algorithm
with $K$ `visible' states, $w_i=h_i$.

It turns out that \textbf{BPI}
achieves a higher capacity than the \textbf{SP} algorithm with $K$
visible states, when $K$ is fixed and $N$ is sufficiently large,
even though the maximal capacity of the binary device is lower.
Interestingly, adding an equivalent of rule R2 to the \textbf{SP}
algorithm allows it to overcome $\textbf{BPI}$.
This issue is further discussed in the Supporting Information.

It is also interesting to note that at very low values of $N$,
performance is better using $20$ states than with an infinite number
of states. Intuitively, this may be due to the fact that in the
unbounded case some synapses are pushed too far and get stuck at high
values of $h_i$, i.e. they lose all their plasticity, while a solution
to the learning problem would require them to come back to the
opposite value of $w_i$.

The last panel in Fig.~\ref{fig:iter_cfr} compares how convergence
time changes with $\alpha$ for the same four algorithms, with the same
number of synapses and same number of states per synapse: while the
cascade model has a clear exponential behavior, the \textbf{BPI} and
\textbf{SP} algorithms maintain nearly constant performance
almost up to their critical point. The \textbf{CP} algorithm is
somehow in between, its performance degrading rapidly with increasing
$\alpha$ (note the logarithmic scale). 

Following the observation that an appropriate number of internal
states $K$ can increase \text{BPI} capacity, we searched for the value
of $K$ that optimizes capacity, and found that it scales roughly as
$\sqrt{N}$ (see the corresponding section and Fig.~8 in the Supporting Information); this is consistent with the
observation that the distribution of hidden states scales as
$\sqrt{N}$ (also discussed in the Supporting Information, see
Fig.~7).  The fact that the capacity is optimal for a finite value of
$K$ makes the \textbf{BPI} algorithm qualitatively different from the
other three, whose performance increases monotonically with $K$.

For a system with a number of states that optimizes
capacity, the optimal value for $p_{s}$ is $0.4$, rather than $0.3$ as
in the unbounded case.  With these settings it is possible to reach a
capacity $\alpha_{c}$ of almost $0.7$ bits per synapse, very close to
the theoretical limit $\alpha_{\max} \simeq 0.83$.  Convergence time at
high values of $\alpha$ scales roughly linearly with $N$, but with a
very small prefactor ($\approx 2 \cdot 10^{-3}$). 


\section{Binary 0,1 neurons and synapses, sparse coding}
\label{sec:01}

\subsection{The 0,1 model neuron}
$\pm 1$ neurons with dense coding (equal probability of $+$ or $-1$
inputs) 
have the biologically unplausible feature of symmetry between the two
states of activity.

A first step towards a more biologically plausible system is to
consider the situation in which both the synaptic weights $w_{j}$ and
neurons are 0,1 binary variables, and the inputs are $\xi_{j}^{a}=1$ with
probability $f$, and $0$ with probability $(1-f)$, where $f \in
[0,0.5]$ is the `coding level'. In this case, we need to take a
non-zero threshold $\theta>0$.  In the following, we choose the
threshold to be $\approx 0.3 N f$ (see Supporting Information
for details).  We consider each input pattern $a$ to have a
desired output $\sigma^{a}=1$ with probability $f$ and $\sigma^{a}=0$
with probability $(1-f)$.

The new classification problem amounts at finding a vector
$\mathbf{w}$ which satisfies the $p$ equations:

\begin{equation}
\sigma^{a}=\Theta \left(\sum_{j=1}^{N}w_{j}\xi_{j}^{a}-\theta\right)\quad\mbox{for $a=1,\ldots,p$}.\end{equation}

\subsection{The optimized algorithm}
The BP scheme can be straightforwardly
applied to the 0,1 perceptron (see Supporting Information
for details); the resulting \textbf{BPI} algorithm is very similar
to the one presented above, with two major differences: \textbf{(i)} 
The quantity to be evaluated at each pattern presentation is not the
total input $I=\sum_{j}\xi_{j}^{\tau}w_{j}^{\tau}$,
but rather the `stability parameter' $\Delta=\left(2 \sigma^{\tau} -
1\right)\left(I-\theta\right)$, which is positive if the
pattern is correctly classified and negative otherwise. \textbf{(ii)} 
Synaptic weights are now computed as
$w_i^{\tau}=\frac{1}{2}\left(\textrm{sign} \left(h_i^{\tau}\right) + 1\right)$,
making the synapse active (inactive) if the hidden variable is positive
(negative), respectively.
The 0,1 algorithm is then the same as the one for the $\pm 1$ case, in which
$\Delta$ replaces $I$.  The performance of this algorithm is
qualitatively very similar to the one for the $\pm 1$ case, with a
lower capacity - about $0.25$, to be compared with a theoretical limit
of $0.59$ \cite{gutfreund90b}.

We have explored variants of the basic \textbf{BPI} algorithm. In
particular, we have studied a stochastic version of the algorithm in
which rule R2 is applied with probability $p_{s}$, but only
for those patterns which require $\sigma^{a}=0$. This \textbf{SBPI01}
algorithm consists in:

\subsubsection*{SBPI01 Algorithm}
{\sf
Compute $I=\xi^{\tau}\cdot w^{\tau}$,
where $w_i^{\tau}=\frac{1}{2}\left(\textrm{sign} \left(h_i^{\tau}\right) + 1\right)$,
and $\Delta=\left(2 \sigma^{\tau} -
1\right) \left(I-\theta\right)$, then 
\begin{enumerate}
\item[(R1)] If $\Delta \ge \theta_m=1$, then do nothing
\item[(R2)] If $0 \le \Delta < \theta_m=1$, then
\begin{enumerate}
\item If $\sigma^{\tau} = 0$, with probability $p_{s}$, if $w_{j}^{\tau}=0$,
then $h_{j}^{\tau+1}=h_{j}^{\tau}-2\xi_{j}^{\tau}$;
\item Else do nothing.
\end{enumerate}
\item[(R3)] If $\Delta < 0$, then
	$h_{i}^{\tau+1}=h_{i}^{\tau}+2\xi_{i}^{\tau}\left(2\sigma^{\tau}-1\right)$.
\end{enumerate}
} 
where  we have introduced $\theta_m$, the threshold for applying rule R2.
Since rule R2 is only applied to patterns with zero output
$\sigma^{\tau}$, the metaplastic changes affect only silent synapses
(for which $w_{j}^{\tau}=0$) involved in the pattern (those for which
$\xi_{i}^{\tau}=1$). Note that using rule R2 only for patterns for
which $\sigma^{a}=0$ not only optimizes performance, but also makes
the algorithm simpler, since in this way there is only the need for
one secondary threshold ($\theta - \theta_m$) instead of two 
(which would have been required if rule R2 had to be applied in all
cases). The opposite choice, i.e. using rule R2 only for patterns
for which $\sigma_{a}=1$, can also be taken with similar results.

As in the preceding case, introducing boundaries for the hidden
variables $h_{j}$ can further improve performance, and the number of
states $K$ which maximizes capacity scales again roughly as $\sqrt{N}$
(shown in the Supporting Information, Fig.~8).
In the case of dense coding, $f=0.5$, and using the optimal value $p_{s}=0.4$,
\textbf{SBPI01} can reach a storage capacity
$\alpha_{c}$ beyond $0.5$ bits per synapse for sufficiently high $N$,
very close to the maximum theoretical value $\alpha_{\max} \simeq 0.59$.

\subsection{Heterogeneous synapses and sparse coding}
One possible way to increase capacity with a very limited number of
available states is to use `sparse' coding, i.e. a low value for $f$.  In an
unsupervised learning scenario, it has been shown that purely binary
synapses (e.g.~only two hidden states) can perform well if $f$ is
chosen to scale as $\log N/N$ \cite{willshaw69,amit94b}. Here, we
chose an intermediate scaling $f=1/\sqrt{N}$. In addition, we also
introduced heterogeneity in synaptic efficacies. Possible synaptic
weights were no longer 0 and 1, but 0 and $a_i$ where $a_{i}$ was
drawn from a Gaussian distribution with mean $1$, and standard
deviation 0.1.  Likewise, the threshold $\theta_m$ used for the
implementation of rule R2 was drawn randomly at each pattern
presentation from a Gaussian distribution centered in $1$ with
variance $0.1$
The resulting algorithm {\bf SBPI-Het} was shown to have very similar
performance to \textbf{SBPI01} in the $f=0.5$ case.

\begin{figure}
\begin{center}
\includegraphics[width=1\columnwidth]{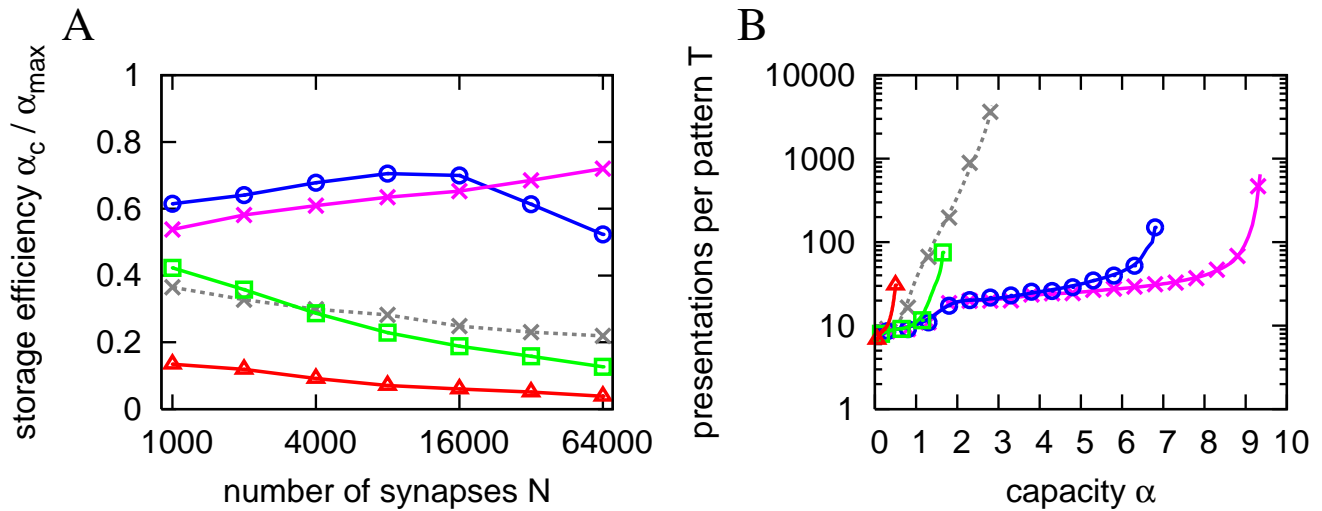}
\end{center}

\caption{Performance of \textbf{SBPI-Het} for different
number of states $K$, with coding level $f=N^{-\frac{1}{2}}$.
Number of samples ranges from $100$ for $N=1000$ to $10$ for $N=64000$.
Triangles:~$K=2$, Squares:~$K=4$, Circles:~$K=10$,
Crosses:~$K=20$. Dashed line:~cascade model with $K=20$.
\textbf{A.}~Storage efficiency vs $N$
\textbf{B.}~Convergence time vs $\alpha$ for $N=64000$
\label{fig:ac_sparse}
}
\end{figure}

In Fig.~\ref{fig:ac_sparse}A we show the maximum capacity $\alpha_{c}$
(defined as for Fig.~\ref{fig:ac_vs_n}) reached in the sparse coding
case divided by the maximum theoretical value $\alpha_{\max}$ (which
depends on $f$), with $p_s=1$, $N$ ranging from $1000$ to $64000$ and
low number of internal states.  The figure shows that a synapse with only two
states (i.e.~with no metaplasticity) has a capacity of only about 10\%
of the maximal capacity in the whole range of $N$ investigated. Adding
hidden states up to $K=10$ improves significantly the performance,
which reaches about 70\% of the maximal capacity for sizes of $N$ of
order 10000. In fact, for such values of $N$ the capacity decreases
when one further increases the number of states. The optimal number of
states increases with $N$ as in the dense coding case, but with a
milder dependence on $N$. In fact, simple arguments based on
unsupervised application of rule R2 predicts in this case an optimal
number of states scaling as $N^{1/4}/\sqrt{\log N}$, which seems to be
roughly consistent with our numerical findings.
Fig.~\ref{fig:ac_sparse}B shows convergence time versus $\alpha$ for
$N=64000$.  It demonstrates again the speed of convergence of the \textbf{SBPI}
algorithm, while the cascade model is significantly slower.

\section{Robustness against noise}
Binary devices have the advantage of simplicity and robustness against noise.
Here we briefly address the issue of resistance against noise which
might affect the multi-stable hidden states.
Intuitively, the fact that the synaptic weights in the \textbf{BPI}
algorithm only depend on the sign of the corresponding hidden
variables, suggests that a device implementing such learning scheme
would be more resistant against accidental changes in the internal
states with respect to a device in which the multi-stable state is
directly involved in the input summation.
We verified this by comparing a perceptron with binary
synapses and $K$ hidden states implementing the \textbf{SBPI}
algorithm with a perceptron with synapses with $K$ visible states
implementing the \textbf{SP} algorithm, both in the unbounded
and in the bounded cases.
The protocols used for testing robustness and the corresponding
results are presented in the Supporting Information.

In all the situations we
tested, we found a pronounced difference between the two devices,
confirming the advantage of using binary synapses in noisy
environments or in presence of unreliable elements.


\section{Discussion}
In this paper, we have shown that simple on-line supervised algorithms
lead to very fast learning of random input-output associations, up to
close to the theoretical capacity, in a system with binary synapses
and a finite number of hidden states. The performance of the
algorithm depends crucially on a rule which leads to synaptic
modifications only if the currently shown pattern is `barely learned'
- that is, a single synaptic flip would lead to an error on that
pattern.  In this situation, the rule requires the synapse to have
metaplastic changes only. Only synapses that contributed to the
correct output need to change their hidden variable, in the direction
of stabilizing the synapse in its current state. This rule originates
directly from the Belief Propagation algorithm.  We have shown that
this addition allows the \textbf{BPI} algorithm to learn a fraction of
bits of information per synapse with at least roughly an order of
magnitude less presentations per pattern than any other known learning
protocol already at moderate system sizes and moderate values of
$\alpha$. Furthermore, for a neuron with about $10^4-10^5$ synapses,
when $\alpha\in[0.3-0.6]$, the \textbf{BPI} algorithm finds a solution
with a few tens of presentations per pattern, while the \textbf{CP}
algorithm is unable to find such a solution in $10^4$ presentations
per pattern. Finally, we showed that this algorithm renders a model
with only two visible synaptic states and $K$ hidden states much more
robust to noise than a model with $K$ visible states.

Other recent studies have considered the problem of learning in
networks with binary synapses. Senn and Fusi \cite{senn05} introduced
a supervised algorithm that is guaranteed to converge for an arbitrary
set of linearly separable patterns, provided there is a finite
separation margin between the two classes. For sets of random
patterns, this last requirement limits learning to a number of
patterns which does not increase with $N$.  Fusi et al \cite{fusi05}
introduced a model that bears similarity with the model we consider
(binary synapses with a finite number of hidden states), with
unsupervised transitions between hidden states. We have shown here
that a supervised version of this algorithm performs significantly
worse than the \textbf{BPI} algorithm.

Since the additional simple rule R2 has such a spectacular effect on
performance, we speculate that neurobiological systems that learn in
presence of supervision must have found a way to implement such a
rule.  The prediction is that when a system learns in presence of an
`error' signal, and synaptic changes occur in presence of that signal,
then metaplastic changes should then occur {\em in absence of the
error signal}, but when the inputs to the system are very close to
threshold. After exposure to such an input, it should be more
difficult to elicit a visible synaptic change, since the synaptic
hidden variables take larger values.

The fact that the algorithms developed here are digital during
retrieval and that discrete (even noisy) hidden variables are only
needed during learning could also have implications in large-scale
electronic implementations, in which the overhead associated with
managing and maintaining multi-stable elements in a reliable way may
be of concern.


\begin{acknowledgments}
This work has been supported by the European Union under the contract
EVERGROW and by Microsoft Research TCI.
\end{acknowledgments}






\begin{thebibliography}{10}
\bibitem{petersen98}
C.~C. Petersen, R.~C. Malenka, R.~A. Nicoll, and J.~J. Hopfield (1998)
\newblock {\em Proc.Natl.Acad.Sci.USA} {\bf 95}, 4732--4737

\bibitem{oconnor05}
D.~H. O'Connor, G.~M. Wittenberg, and S.~S-H. Wang (2005)
\newblock {\em Proc Natl Acad Sci U S A} {\bf 102}, 9679--9684

\bibitem{lisman85}
J.~E. Lisman (1985)
\newblock {\em P.~N.~A.~S.~USA} {\bf 82}, 3055--3057

\bibitem{zhabotinsky00}
A.~M. Zhabotinsky (2000)
\newblock {\em Biophys.~J.} {\bf 79}, 2211--2221

\bibitem{bhalla99}
U.~S. Bhalla and R.~Iyengar (1999)
\newblock {\em Science} {\bf 283}, 381--387

\bibitem{willshaw69}
D.~Willshaw, O.~P. Buneman, and H.~Longuet-Higgins (1969)
\newblock {\em Nature} {\bf 222}, 960--962

\bibitem{marr69}
D.~Marr (1969)
\newblock {\em J.~Physiol.} {\bf 202}, 437--470

\bibitem{sompolinsky86}
H.~Sompolinsky (1986)
\newblock {\em Phys. Rev. A} {\bf 34}, 2571--2574

\bibitem{amit92b}
D.~J. Amit and S.~Fusi (1992)
\newblock {\em Network} {\bf 3}, 443--464

\bibitem{amit94b}
D.~J. Amit and S.~Fusi (1994)
\newblock {\em Neural Computation} {\bf 6}, 957--982

\bibitem{brunel98b}
N.~Brunel, F.~Carusi, and S.~Fusi (1998)
\newblock {\em Network} {\bf 9}, 123--152

\bibitem{fusi05}
S.~Fusi, P.~J. Drew, and L.~F. Abbott (2005)
\newblock {\em Neuron} {\bf 45}, 599--611

\bibitem{fusisenn}
S. Fusi, W. Senn (2005)
\newblock {\em Physical Review E} {\bf 71}, 061907
				   
\bibitem{tsodyks90}
M.~Tsodyks (1990)
\newblock {\em Mod.~Phys.~Lett.~B} {\bf 4}, 713--716

\bibitem{rosenblatt62}
F.~Rosenblatt (1962)
\newblock {\em Principles of neurodynamics}.
\newblock Spartan Books, New York, 1962.

\bibitem{minsky69}
M.~Minsky and S.~Papert (1969)
\newblock {\em Perceptrons: An Introduction to Computational Geometry}.
\newblock MIT Press, Cambridge, Ma

\bibitem{kennedy00}
M.~B.~Kennedy (2000)
\newblock {\em Science} {\bf 290}, 750-754

\bibitem{krauth89}
W.~Krauth and M.~M\'ezard (1989)
\newblock {\em J.~Phys.~France} {\bf 50}, 3057--3063

\bibitem{krauth89b}
W.~Krauth and M.~Opper (1989)
\newblock {\em J. Phys. A} {\bf 22}, L519-L523

\bibitem{senn05}
W.~Senn and S.~Fusi (2005)
\newblock {\em Physical Review E} {\bf 71}, 061907

\bibitem{gardner88}
E.~J.~Gardner (1988)
\newblock {\em  J. Phys. A} {\bf 21}, 257-270

\bibitem{NPC-learning1}
R.L.~Rivest A.L.~Blum (1992)
\newblock {\em Neural Networks} {\bf 5}, 117--127

\bibitem{NPC-learning2}
A.~Amaldi (1991)
\newblock In O.~Simula T.~Kohonen, K.~Makisara and J.~Kangas, editors, {\em
  Artificial Neural Networks}, vol.~1, 55--60. Elsevier Science
  Publisher, North-Holland, Amsterdam

\bibitem{MPZ}
G.~Parisi, M.~Mezard, and R.~Zecchina (2002)
\newblock {\em Science} {\bf 297}, 812-815

\bibitem{MZ}
M.~Mezard and R.~Zecchina (2002)
\newblock {\em Phys. Rev. E} {\bf 66}, 056126

\bibitem{BMZ}
M.~Mezard, A.~Braunstein and R.~Zecchina (2005)
\newblock {\em Random Structures and Algorithms} {\bf 27}, 201--226

\bibitem{COLORING}
A.~Braunstein, R.~Mulet, A.~Pagnani, M.~Weigt, and R.~Zecchina (2003)
\newblock {\em Phys. Rev. E} {\bf 68}, 036702

\bibitem{nosotros}
A.~Braunstein and R.~Zecchina (2006)
\newblock {\em Phys. Rev. Lett.} {\bf 96}, 030201

\bibitem{amit89}
D.~J. Amit, C.~Campbell and K.~Y.~M.~Wong (1989)
\newblock {\em J. Phys. A.: Math. Gen.} {\bf 22}, 4687-4693

\bibitem{gutfreund90b}
H.~Gutfreund and Y.~Stein (1990)
\newblock {\em J.~Phys.~A: Math.~Gen.} {\bf 23}, 2613--2630


\end{thebibliography}
\end{document}